\documentclass{aa}
\usepackage{graphicx}
\usepackage{natbib}
\bibpunct{(}{)}{;}{a}{}{,}
\begin{document}
\title{Detection of differential rotation in $\psi$ Cap with profile
  analysis\thanks{Based on observations collected at the European
    Southern Observatory, La Silla (65.L-0101)}}

   \author{A. Reiners\inst{1}
          \and
          J.H.M.M. Schmitt\inst{1}
          \and
          M. K\"urster\inst{2}
          }

   \offprints{A. Reiners}

   \institute{Hamburger Sternwarte, Universit\"at Hamburg,
     Gojenbergsweg 112, 21029 Hamburg, Germany\\
     \email{areiners@hs.uni-hamburg.de}
     \and
     European Southern Observatory, Casilla 19001, Vitacura, Santiago 19, Chile}

   \date{}

   \abstract{We report detection of differential rotation on the F5
     dwarf $\psi$\,Cap using line profile analysis. The Fourier
     transform of both Fe\,\textsc{i}\,$\lambda$5775 and
     Si\,\textsc{i}\,$\lambda$5772 are used to obtain a projected
     rotational velocity of $v\,\sin{i}\,=\,42\,\pm\,1$\,km\,s$^{-1}$.
     Modelling of the Fourier transformed profiles shows that the
     combined effects of equatorial velocity, inclination and
     differential rotation dominate the line profile while limb
     darkening and turbulence velocities have only minor effects.
     Rigid rotation is shown to be inconsistent with the measured
     profiles.  Modelling the line profiles analogous to solar
     differential rotation we find a differential rotation parameter
     of $\alpha\,=\,0.15\,\pm\,0.1$ ($15\,\pm\,10$\%) comparable to
     the solar case. To our knowledge this is the first successful
     measurement of differential rotation through line profile
     analysis.   }

   \maketitle
%

\section{Introduction}
\label{Introduction}

According to the standard paradigm of stellar activity differential
rotation is a central ingredient for the magnetic dynamo presumed to
underlie all activity phenomena. In a turbulent convection zone
differential rotation is expected to be a central element of stellar
activity. The interaction of rotation and convection naturally
produces deviations from rigid rotation. One example for this effect
is the Sun, whose equator rotates $\sim$\,20\% faster than higher
latitudes \citep[e.g.][]{Lang92}. Direct predictions of the dependence of
differential rotation on rotational velocity and spectral type have
been made \citep[e.g.][]{Belvedere80,Kitchatinov99}.  Observations
suggest a strong correlation between rotation and activity in
late-type stars with many pieces of evidence summarized as
``rotation-activity connection'' \citep{Hempelmann95, Messina01}.

However, measurements of differential rotation among stars are either
complicated or time-consuming or both. One method to determine
differential rotation is to examine spectral line shapes. Subtle
differences exist between the line profile of a rigidly rotating star
and that of a differentially rotating star, because the underlying
flow fields are different. Although significant differences between
the two cases are predicted, all previous attempts to measure
differential rotation through line profiles alone remained
unsuccessful and yielded results consistent with rigid rotation
\citep[e.g.  ][]{Gray82,Dravins90,Groot96}. At any rate, for a
successful measurement one needs high spectral resolution and high
signal to noise. With the Very Long Camera (VLC) of the Coud\'e
Echelle Spectrograph (CES) at ESO's 3.6m telescope an instrumental
setup satisfying these requirements is available. In this letter we
report the detection of differential rotation in the F5 dwarf
$\psi$\,Cap, to our knowledge the first successful measurement of
differential rotation through line profiles.

\section{Data}
\label{Data}

$\psi$\,Cap (HD\,197\,692, F5\,V, $V_{\rm mag}\,=\,4.13$,
$v\,\sin{i}\,=\,40$\,km\,s$^{-1}$; \citet{Uesugi82}) is the fastest
rotator in a set of solar-like stars in the solar neighbourhood we
observed on October 13, 2000 at ESO's 3.6m telescope (La Silla). The
spectral resolution achieved with the CES / VLC setup was
$R\,=\,235\,000$ ($\sim$\,1.28\,km\,s$^{-1}$). Three consecutive
exposures of $\psi$\,Cap of 270\,s each covering the wavelength range
between 5770 -- 5810\,\AA \ were taken. The signal to noise ratio of
the extracted spectrum is $S/N \sim 800$ per pixel with proper flat
fielding and removal of interference pattern \citep[c.f.][
Sect.\,6]{Kürster01}.

We concentrate our analysis on Fourier transforms of
well isolated lines (Fe\,\textsc{i}\,$\lambda$5775 and 
Si\,\textsc{i}\,$\lambda$5772),
for which continuum placement because of line blending
is not a problem even for large $v\,\sin{i}$ values.  
The signal to noise ratio of our data
is high enough to show the crucial features in the Fourier transform
already from a single line.

\begin{figure}
  \centering \resizebox{.8\hsize}{\hsize}{\includegraphics[]{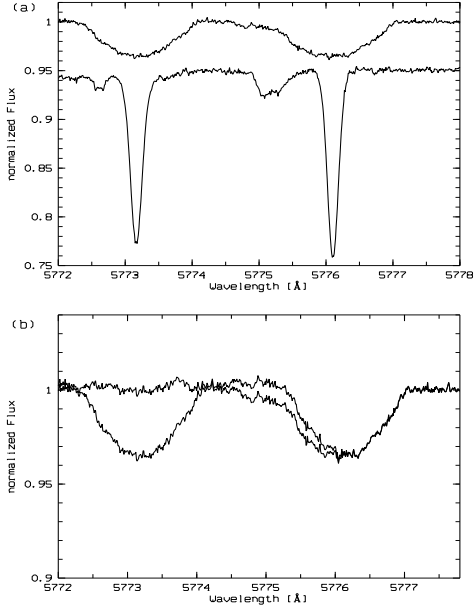}}
  \caption{(a) Fe\,\textsc{i}\,$\lambda$5775 (right) and 
    Si\,\textsc{i}\,$\lambda$5772 (left) lines of $\psi$\,Cap (upper
    line) and $\iota$\,Psc (lower line; shifted along y-axis). (b)
    Original and modified line profiles of $\psi$\,Cap (see text).}
\label{spectrum} 
\end{figure}

Still, in Fe\,\textsc{i}\,$\lambda$5775 a small blend occurs from a
neighbouring set of weak lines $\sim\,0.8$\,\AA\ apart, the influence
of which on differential rotation determination is not clear.  During
our observing run a spectrum of the slow rotator $\iota$\,Psc
(HD\,222\,368; F7\,V) was taken. In Fig.\ref{spectrum}a both spectra
are shown, with the profile of $\iota$\,Psc shifted according to its
relative radial velocity. Deblending of Fe\,\textsc{i}\,$\lambda$5775
was accomplished by, first, removing the line in the spectrum of
$\iota$\,Psc ``by hand'', broadening this template with $v_{\rm
  rot}\,=\,40$\,km\,s$^{-1}$ and subtracting it from the $\psi$\,Cap
spectrum. In Fig.~\ref{spectrum}b the recorded spectrum as well as the
modified spectrum of $\psi$\,Cap are shown. Obviously, the
Si\,\textsc{i}$\lambda$5772 equivalent widths of $\psi$\,Cap and
$\iota$\,Psc are the same, since it has been successfully removed (cf.
Fig.\ref{spectrum}b). The Fe\,\textsc{i}\,$\lambda$5775 line appears
symmetric, thus we think that deblending was successful.

\section{Method of Analysis}
\label{Method}

Absorption line profiles are influenced by a number of different
effects. An absorption line profile at any point on the star is
determined by temperature, gravity, element abundances and atomic
parameters. This ``intrinsic'' profile is Doppler broadened by
velocity fields. Many efforts have been undertaken to distinguish
these velocity fields, one of them being the stellar rotation
\citep[see][ and references therein]{Gray88}. In addition to the
projected rotational velocity of the star, radial-tangential macro-
and isotropic microturbulence (denoted with $\zeta_{\rm RT}$ and $\xi$
resp.)  turned out to be a reasonable parameterization of stellar
atmospheric velocity fields. These parameterizations assume that
Doppler broadenings can be treated as convolutions; and that the
``intrinsic'' profiles are identical over the apparent stellar disk.
For fast rotators ($v\,\sin{i}\,>\,30$\,km~s$^{-1}$) and
stationary atmospheres rotational broadening dominates and no
complications occur with this assumption.

Interpreting the observed data profile $D(\lambda)$ as a a multiple
convolution (denoted by $\ast$) between the intrinsic profile
$F(\lambda)$ (including microturbulence broadening), the rotational
broadening profile $G(\lambda)$, the instrumental profile $I(\lambda)$
and the macroturbulence profile $M(\lambda)$, $D(\lambda)$ can be
written as
\begin{equation}
  D(\lambda) = F(\lambda) \ast G(\lambda) \ast I(\lambda) \ast M(\lambda).
  \label{convolution}
\end{equation}

We calculated $F(\lambda)$ using the packages \textsc{Atlas9}
\citep{Kurucz79, Kurucz93} and \textsc{Bht} \citep{Baschek}. Atomic
damping coefficients obtained from VALD \citep{VALD, Kurucz94} were
included in the profile calculations; we chose solar metallicity and
for the velocity dispersion of the opacity distribution function in
\textsc{Atlas9} we used 1.0\,km~s$^{-1}$. The temperature has been set
to $T_{\rm eff}\,=\,6500$\,K \citep{Blackwell98}. To calculate
$G(\lambda)$ a modified version of a package developed and described
by \citet{Townsend97} is used. The surface integration is carried out
over about 25\,500 visible surface elements. The adopted
limb-darkening law is given by
\begin{equation}
  I(\cos{\theta})\,=\,I_{0}\,(1 - \epsilon + \epsilon \cos{\theta}),
\end{equation}
with $\theta$ denoting the angle between the surface element normal
and the observer's line of sight and $\epsilon$ the limb darkening
coefficient. We parametrize the differential rotation law through
\begin{equation}
\label{difflaw}
  \omega(l)\,=\,\omega_{0} - \omega_{1} \sin^2{l}
\end{equation}
with $l$ being the latitude. Specifically, differential rotation is
expressed in terms of $\alpha\,=\,\omega_{1}/\omega_{0}$.  The
differential rotation law (Eq.\,\ref{difflaw}) is adopted from the
solar case, where $\alpha_{\sun}\,=\,0.20$; a more general approach
would expand $\omega(l)$ in terms of orthogonal polynomials. The
instrumental profile $I(\lambda)$ has been determined from the shapes
of calibration lamp lines, a correction for thermal broadening
according to the lamp temperature of $120\degr$C was applied. The
macroturbulence broadening function $M(\lambda)$ is adopted from
\citet[ p.~407]{Gray92}.

The contributions of different velocity fields are difficult to
separate especially in the wavelength domain. In the Fourier domain
convolutions become multiplications which are much easier to handle;
the advantages of Fourier domain are discussed in detail by
\citet{Gray92}. In the Fourier domain Eq.~\ref{convolution} becomes
\begin{equation}
  d(\sigma) = f(\sigma) \cdot g(\sigma) \cdot i(\sigma) \cdot m(\sigma) .
  \label{convolution-fourier}
\end{equation}
The Fourier frequency $\sigma$ is expressed in
cycles~/~(km~s$^{-1}$)~=~s~km$^{-1}$. Noise in Fourier domain can be
expressed as
$S_{\sigma}\,=\,S_{\lambda}\,\Delta\lambda\,N^{\frac{1}{2}}$
\citep{Gray92}, with $S_{\lambda}$ being data noise and
$\Delta\lambda$ step size in the wavelength domain, N is the number of
data points. The Fourier transform of a real, symmetric profile yields
a real function in the Fourier domain. Investigations of imaginary
transforms of asymmetric lines show that these are no reliable tracers
of profile properties \citep{Gray80}. We created a symmetric profile
by mirroring the absorption profile at it's center; the achieved
Fourier transform turned out to be stable against small variations of
the center position. Rotational broadening, which dominates the line
broadening in the case of $\psi$\,Cap, yields symmetric profiles.
Asymmetries due to convection are believed to be of the order
$v\,<\,1\,$km\,s$^{-1}$.  In Fourier space this is
$\sigma\,>\,1\,$s\,km$^{-1}$ while our analysis focuses on
$\sigma\,<\,0.04\,$s\,km$^{-1}$. In the case of
Fe\,\textsc{i}\,$\lambda$5775 an asymmetry may occur due to imperfect
deblending, for Si\,\textsc{i}\,$\lambda$5772 no mechanism should
contribute asymmetries of this order.

The Fourier transformed observed line profile was compared with a
Fourier transformed model profile via a $\chi^{2}$ test.
Within our adopted modelling
approach the following six fit parameters determine the model
profiles: rotational velocity ($v$), inclination angle ($i$),
differential rotation ($\alpha$), limb darkening ($\epsilon$), macro-
($\zeta_{RT}$) and microturbulence ($\xi$).  The
$\chi^{2}$-calculations have always been carried out directly on
$d(\sigma)$. No further corrections have been applied to the data to
avoid complications with amplified noise. \citet{Bruning84} pointed
out that the convolution method induces systematic errors due to
incorrectly estimated line depths especially in slow rotators. Our
study is not focused on reproducing equivalent widths, which can be
tuned e.g. by element abundances. The main goal is to reproduce the
line shapes. Although we do not examine slow rotators in this study,
normalized transformed profiles were used, that is, Fourier
transformed profiles have been scaled to unity at $\sigma\,=\,0$.

\section{Results and Discussion}
\label{Results}

After a rough determination of $v\,\sin{i}$ a grid calculation in the
six fit parameters was carried out. As to our model parameterization
we note that this description contains parameters whose physics is
poorly understood, i.e., the micro- and macroturbulence. We thus use
$\chi^{2}$ only as a relative parameter and not as an absolute
criterion as to whether a model is acceptable or not; $\chi^{2}_{\rm
  n}$ denotes the goodness of fit relative to the best fit, for which
$\chi^{2}_{\rm n}\,=\,1$ is set. For the actual grid we chose the
parameter ranges of $\epsilon$ (0.4 -- 0.8), $\zeta_{\rm RT}$ (4.0 --
7.0)\,km\,s$^{-1}$ and $\xi$ (1.0 -- 2.5)\,km\,s$^{-1}$. The rotation
velocity was limited to $v\,<\,120$\,km\,s$^{-1}$ ($i\,>\,20\degr$).

Goodness of fit calculations have been carried out on
Fe\,\textsc{i}\,$\lambda$5775 and Si\,\textsc{i}\,$\lambda$5772; both
lines show identical results. In the following we show only the
results for Fe\,\textsc{i}\,$\lambda$5775.

\begin{figure}
  \centering
  \resizebox{.9\hsize}{!}{\includegraphics[angle=-90]{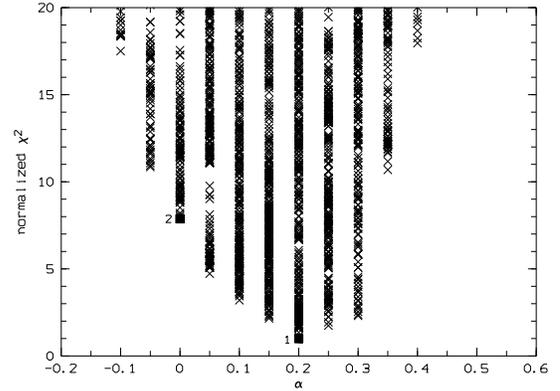}}
  \caption{Normalized $\chi^{2}$ vs. $\alpha$ for calculated models 
    with various values of $v, i, \epsilon, \zeta_{\rm RT}$ and $\xi$
    (see text); each cross represents one set of parameters.  Best
    fits of all models (1) and for rigid rotation (2) are denoted by
    full squares.}
\label{envelope}
\end{figure} 
In Fig.\,\ref{envelope} we plot the resulting $\chi^{2}_{\rm n}$ as a
function of the differential rotation parameter $\alpha$; for each
fixed value of $\alpha$, we plot the $\chi^{2}_{\rm n}$ values
obtained by varying all other model parameters. Clearly, a well
defined lower envelope curve exists with a minimum at
$\alpha\,=\,0.20$. The best fits for differential rotation (here
$\alpha\,=\,0.20$; $\chi^{2}_{\rm n}\,=\,1$, Model\,1) and rigid
rotation ($\alpha\,=\,0.0$, $\chi^{2}_{\rm n}\,= 7.9$, Model\,2) are
marked by full squares; all parameters are shown in
Tab.\,\ref{Parameters}. 

\begin{figure}
  \centering
  \resizebox{.9\hsize}{!}{\includegraphics[angle=-90]{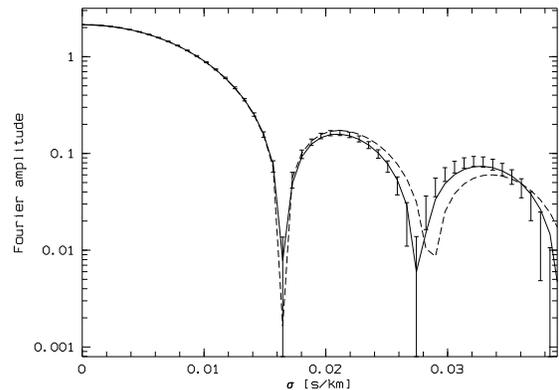}}
  \caption{The Fourier transformed data indicated by error bars. 
    Best fits for differential rotation ($\alpha\,=\,0.20$; full line,
    Model\,1) and rigid rotation ($\alpha\,=\,0.0$; dashed line,
    Model\,2) are shown. See Tab.\,\ref{Parameters} for parameters.}
\label{FourierPlot} 
\end{figure}

\begin{table}
  \caption[]{Parameters of best fitting models for rigid and differential rotation}
  \label{Parameters}
  $$
  \begin{array}{cp{0.2cm}|p{0.2cm}ccccccr}
    \hline
    \noalign{\smallskip}
    $Model$ & & & v & i & \alpha & \epsilon & \zeta_{\rm RT} & \xi & \chi^{2}_{n} \\
    \noalign{\smallskip}
    \hline
    \noalign{\smallskip}
    1 & & & 43\frac{\rm km}{\rm s} & 80^{\circ} & 0.20 & 0.6 & 4.0\frac{\rm km}{\rm s} & 1.0\frac{\rm km}{\rm s} & 1.0 \\
    2 & & & 48\frac{\rm km}{\rm s} & 60^{\circ} & 0.00 & 0.8 & 7.0\frac{\rm km}{\rm s} & 2.5\frac{\rm km}{\rm s} & 7.9 \\
    \noalign{\smallskip}
    \hline
  \end{array}
  $$
\end{table}
In Fig.\,\ref{FourierPlot} we plot the corresponding profiles in the
Fourier domain; obviously, Model\,1 provides a much better fit than
Model\,2, which is not an adequate description of our data.  While no
statistical uncertainties can be given on our $\chi^{2}_{\rm
  n}$-values, Fig.\,\ref{FourierPlot} shows $\chi^{2}_{\rm n}$-values
of 7.9 to be clearly unacceptable.

Inspection of the calculated $\chi^{2}_{\rm n}$-grid shows that the
parameters $\epsilon, \zeta_{\rm RT}$ and $\xi$ produce only second
order effects; the fits are driven by the chosen values of $v$, $i$
and $\alpha$, and we cannot determine all model parameters
independently. Therefore, we fixed the values $\epsilon\,=\,0.6$
\citep{Carbon69}, $\zeta_{\rm RT}\,=\,6.0$\,km\,s$^{-1}$ and
$\xi\,=\,1.0$\,km\,s$^{-1}$ \citep[ and references therein]{Gray88}.
We emphasize that our subsequent findings do not require these
specific parameter settings, and that the influences of $\epsilon$,
$\zeta_{\rm RT}$ and $\xi$ are too small to revoke the effects of
$\alpha$, $v$ and $i$. For the given parameters the best fit value for
$v\,\sin{i}$ is 42.3\,km\,s$^{-1}$.  We estimate the total systematic
errors to be $\approx\,1$\,km\,s$^{-1}$. Models outside the range
$42.3\,\pm\,1$\,km\,s$^{-1}$ all have $\chi^{2}_{\rm n}\,>\,5$.

\begin{figure}
  \centering
  \resizebox{.9\hsize}{!}{\includegraphics[angle=-90]{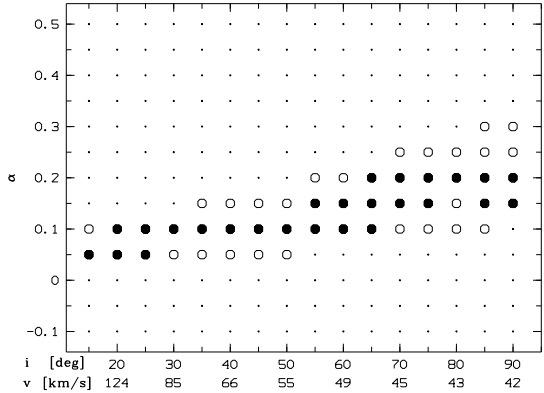}}
  \caption{$\chi^{2}_{\rm n}$ of fits to  Fe\,\textsc{i}\,$\lambda$5775 with 
    41.3\,km\,s$^{-1} < v\,\sin{i} < 43.3$\,km\,s$^{-1}$,
    $\epsilon\,=\,0.6$, $\zeta_{\rm RT}\,=\,6.0$\,km\,s$^{-1}$ and
    $\xi\,=\,1.0$\,km\,s$^{-1}$ in the $\alpha$\,--\,$i$ ($v$) plane:
    $\chi^{2}_{\rm n} < 5 \,(\bullet); \,5 < \chi^{2}_{\rm n} < 10
    \,(\circ); \chi^{2}_{\rm n} > 10 \,(\cdot)$.}
\label{contour} 
\end{figure}

Theoretical investigations of the line profile behaviour in Fourier
space show that with differential rotation $\alpha\,\neq\,0$ the
inclination $i$ becomes important \citep[e.g.][]{Bruning81}. For
comparable values of $v\,\sin{i}$ the best fit values of $\alpha$ and
$i$ are correlated.  In Fig.~\ref{contour} we consider calculated
models with fixed $\epsilon$, $\zeta_{\rm RT}$, $\xi$, $v\,\sin{i}$
between 41.3 and 43.3\,km\,s$^{-1}$ and varying $\alpha$, $v$ and $i$.
Three groups of models are distinguished; $\chi^{2}_{\rm n}\,<\,5
\left( \bullet \right)$, $5\,<\chi^{2}_{\rm n}\,<\,10 \left( \circ
\right)$ and $\chi^{2}_{\rm n}\,>\,10 \left(\,\cdot\,\right)$.
A well defined area of reliable fits in the $\alpha$\,--\,$i$ plot
emerges. All fits with $\chi^{2}_{\rm n}\,<\,10$ show
$\alpha\,>\,0.0$. Although we used no absolute criterion on wheter a
model is acceptable, the estimation of systematic errors and a
comparison with Fig.\,\ref{FourierPlot} shows that the threshold
$\chi^{2}_{\rm n}\,<\,10$ is a rather high choice.

A variety of combinations of equatorial velocity and inclination seems
possible for $\alpha\,>\,0$. A smaller differential effect is preferred
for faster rotating models.  For extremely high values of $v >
200$\,km\,s$^{-1}$ rigid rotation might be argued, but such velocities
seem unlikely given the age of $\psi$\,Cap of $\sim\,2\,10^{9}$\,a
\citep{Lachaume99}. Even the minimum rotational velocity of
42\,km\,s$^{-1}$ seems uncommonly high. From Fig.\,\ref{contour} we
derive $\alpha\,=\,0.15\,\pm\,0.1$. The identical result was found for
Si\,\textsc{i}\,$\lambda$5772, where the same procedure was applied.

To summarize differential rotation has been established for the rapid
rotator $\psi$\,Cap independently from two absorption line profiles.
While $\psi$\,Cap rotates at least 20 times faster than the Sun, it's
differential rotation is comparable to the solar value, but not with
the differential rotation patterns determined from Doppler images of
the fast rotators AB\,Dor \citep{Donati97} and PZ\,Tel
\citep{Barnes00}. As direct predictions for a F5 dwarf have not been
calculated by \citet{Kitchatinov99} and the rotation period of
$\psi$\,Cap is only poorly determined, the consistency of our result
and the model is not clear. Assuming $R\,=\,1.2\rm{\,R_{\sun}}$ and
$i\,=\,90\degr$, we find for $\psi$\,Cap the rotation law
$\omega(l)\,=\,4.38 - 0.66\,\sin^{2}{l}$\,rad/d; $\psi$\,Cap does not
rotate like a rigid body as suggested for AB\,Dor and PZ\,Tel. To
compare theory and observations more detailed predictions especially
on spectral type dependence and a greater sample of direct
observations are needed. In particular verification of the
differential rotation results of line profile analysis and Doppler
imaging for the same star will be instructive.

\begin{acknowledgements}
  A.R. acknowledges financial support from Deutsche
  Forschungsgemeinschaft DFG-SCHM 1032/10-1.
\end{acknowledgements}

\end{document}